\documentclass[%
	twocolumn,
        aps,pra,showpacs,amsmath,amssymb]{revtex4}

\pdfoutput=1
\usepackage{hyperref}
\hypersetup{pdftitle={Cheon's anholonomies in Floquet operators},%
	pdfauthor={Manabu Miyamoto and Atushi Tanaka}}
\usepackage{graphicx}

\newcommand{\bra}[1]{\langle{}#1{}|}
\newcommand{\ket}[1]{|{}#1{}\rangle}
\newcommand{\bracket}[2]{\langle{}#1{}|{}#2{}\rangle}
\newcommand{\braOket}[3]{\langle{}#1{}|{}#2{}|{}#3{}\rangle}
\newcommand{\ketbra}[2]{|{}#1{}\rangle\langle{}#2{}|}

\newcommand{\pdfrac}[2]{\frac{\partial #1}{\partial #2}}
\newcommand{\set}[1]{\left\{ #1 \right\}}
\newcommand{\bbset}[1]{\mathbb{#1}}
\newcommand{\Integer}{\bbset{Z}}
\newcommand{\Hilbert}{\mathcal{H}}

\newcommand{\Torder}{\mathop{\mathcal{T}}_{\leftarrow}}       
\newcommand{\up}{\uparrow}
\newcommand{\down}{\downarrow}
\newcommand{\bvec}[1]{\boldsymbol{#1}} 

\newcommand{\lambdai}{\lambda_{\rm i}}
\newcommand{\lambdaf}{\lambda_{\rm f}}
\newcommand{\Nt}{M}

\begin{document}

\title{Cheon's anholonomies in Floquet operators}

\author{Manabu Miyamoto}
\email{miyamo@hep.phys.waseda.ac.jp}
\affiliation{Department of Physics, Waseda University, 
  Okubo, Shinjuku-ku, Tokyo 169-8555, Japan}
\author{Atushi Tanaka}
\email{tanaka@phys.metro-u.ac.jp}
\affiliation{Department of Physics, Tokyo Metropolitan University,
  Minami-Osawa, Hachioji, Tokyo 192-0397, Japan}


\begin{abstract}
Anholonomies in the parametric dependences of the eigenvalues and 
the eigenvectors of Floquet operators that describe unit time 
evolutions of periodically driven systems, 
e.g., kicked rotors, are studied.
First, an example of the anholonomies induced by a periodically pulsed 
rank-1 perturbation is given.
As a function of the strength of the perturbation,
the perturbed Floquet operator of the quantum map and its spectrum are 
shown to have a period.
However, we show examples where each eigenvalue does not obey
the periodicity of the perturbed Floquet operator
and exhibits an anholonomy.
Furthermore, this induces another anholonomy in the eigenspaces,
i.e., the directions of the eigenvectors, of the Floquet operator.
These two anholonomies are previously observed in a family of
Hamiltonians [T.~Cheon, Phys.~Lett.~A {\bf 248}, 285 (1998)]
and are different from the phase anholonomy known as geometric phases.
Second, the stability of Cheon's anholonomies in periodically driven 
systems is established by a geometrical analysis of the family of 
Floquet operators.
Accordingly, Cheon's anholonomies are expected to be abundant in systems 
whose time evolutions are described by Floquet operators.
As an application, a design principle for quantum state manipulations 
along adiabatic passages is explained.
\end{abstract}

\pacs{03.65.Vf}

\maketitle

\section{Introduction}
\label{sec:introduction}
The parametric dependence of an eigenvector of an operator
often exhibits anholonomy in its phase
~\cite{GeometricPhaseReview}.
A simple demonstration of the phase anholonomy in an eigenvector
of a Hamiltonian is shown by Berry~\cite{Berry:PRSLA-430-405}:
Prepare the system to be in an eigenstate of
the Hamiltonian, whose energy spectrum is assumed to be discrete and
nondegenerate.
During the adiabatic change of the parameters of the Hamiltonian,
which is kept to be nondegenerate along the change, 
the system continuously remains to be in an 
eigenstate of the instantaneous Hamiltonian, according to 
the adiabatic theorem~\cite{Born:ZP-51-165}.
When the parameter returns to its initial value, 
after the adiabatic change along a closed path in the parameter space,
the difference between the initial and the final state vectors is only 
in its phase, which is composed by two ingredients: One is
called a dynamical phase that is determined by the accumulation of the 
eigenenergy along the adiabatic time evolution. The other is called 
a geometric phase, or the phase anholonomy that reflects 
the geometric 
structure of the family of eigenvectors in the parameter space.
There is a non-Abelian generalization of the phase anholonomy:
This was pointed out by Wilczek and Zee in the parametric change of an
eigenspace of a Hamiltonian that has a spectral degeneracy%
~\cite{Wilczek:PRL-52-2111}.
The phase anholonomy appears in various fields of physics, besides
quantum mechanics, and brings profound 
consequences~\cite{GeometricPhaseReview}.

Recently, Cheon found exotic anholonomies, which are completely different
from the conventional phase anholonomy, in a family of systems with 
generalized pointlike potentials~\cite{Cheon:PLA-248-285}.
Cheon's anholonomies appear, surprisingly, both in eigenenergies and
eigenvectors: 
The trail of an eigenenergy along a change of parameters 
on a closed path that encircles a singularity does not draw a closed curve 
but, instead, a spiral. 
Since the initial and the final eigenenergies in the closed path are 
different eigenvalues of a Hermite operator,
the corresponding eigenvectors must be orthogonal.
Hence the eigenenergy anholonomy induces another anholonomy in the
direction of eigenvectors.
The origin of Cheon's anholonomies in the family of systems with the 
generalized pointlike potentials
is identified with the geometrical structure of the family's 
parameter space~\cite{Cheon:AP-294-1,Tsutsui:JMP-42-5687}.

In order to distinguish Cheon's anholonomy in the directions of eigenvectors
from 
Wilczek-Zee's phase anholonomy, which
requires a degenerate spectrum and
transports an eigenvector into its nonorthogonal 
direction in general along adiabatic changes on closed paths,
we will call the former an {\em eigenspace anholonomy\/}:
Wilczek-Zee's phase anholonomy concerns the change of an eigenvector within 
a single and degenerate eigenspace and 
Cheon's eigenspace anholonomy, which do not require spectral degeneracies,
concerns the journey of an eigenvector from one eigenspace 
into another eigenspace.

We can easily expect that Cheon's anholonomies
would bring profound consequences in various fields of physics,
as is done by the phase anholonomy.
For example, in the adiabatic (sometimes referred to as 
Born-Oppenheimer~\cite{Born:AP-84-457}) 
approximation~\cite{Born:1954}, 
it has been considered to be legitimate to assume that 
an adiabatic potential surface, which is an eigenvalue of an electronic
Hamiltonian with a frozen nuclear configuration, is a single-valued 
function of the nuclear configuration. The single-valuedness would 
be broken if Cheon's eigenenergy anholonomy emerged. A similar question
may be raised in the Bloch theory in solid state physics~\cite{Kittel:ISSP}.
At the same time, Cheon's anholonomies may be applied to manipulate
quantum systems to transfer a quantum state adiabatically into
another state, as is suggested by Cheon~\cite{Cheon:PLA-248-285}.
The last point will be discussed more precisely in this paper.
However, all known examples of the eigenenergy anholonomy, up to now,
require an exotic connection condition 
around a singular potential~\cite{Tsutui:JPA-36-275}.
Hence it is still worth to find systems that exhibit 
Cheon's anholonomies.

The purpose of the present paper is to show Cheon's anholonomies
in periodically driven systems.
More precisely, we will discuss
quasienergy and eigenspace anholonomies with respect to Floquet operators 
that describe unit time evolutions of the periodically driven systems.
First, we provide an instance of a quantum map, i.e., a quantum system 
under a periodically pulsed perturbation~\cite{QuantumMap}.
The simplicity of the Floquet operators of quantum maps
allows us a thorough analysis.
In order to prepare it, the parametric dependence,
induced by the change of the strength of the perturbation, 
of eigenvectors of the Floquet operators of quantum maps is reviewed in 
Section~\ref{sec:adiabaticTransport}.
In Section~\ref{sec:rank1}, a quantum map that is perturbed by
a rank-$1$ operator is introduced.
The details of its properties are explained 
in Appendices~\ref{sec:normalForm} and~\ref{sec:cyclicity}.
In Section~\ref{sec:existance}, it is 
shown that the rank-$1$ perturbation, with respect to the original 
Floquet operator, enables us to introduce a family of Floquet 
operators to realize Cheon's anholonomies.
Several examples are shown in Section\ref{sec:examples}.
Second, the stability of the anholonomies is examined.
A geometrical analysis, which is shown in Section~\ref{sec:abundance},
of the family of Floquet operators
elucidates that the appearance of Cheon's anholonomies is not
restricted in the periodically pulsed systems and is also possible
in periodically driven systems in general.
Furthermore, we may claim that Cheon's anholonomies are abundant 
in systems whose time evolutions are described by Floquet operators.
Among possible consequences and applications of our result,
Section~\ref{sec:manipulation} provides a discussion on a design 
principle on quantum state manipulations along adiabatic passages. 
Section~\ref{sec:summary} provides a discussion and a summary.
A part of the present result was briefly announced in Ref.~\cite{TM061}.

\section{Adiabatic transport of eigenvectors in a quantum map}
\label{sec:adiabaticTransport}

To prepare our analysis of quantum maps,
we review the parametric motions of eigenvectors and eigenvalues of 
Floquet operators and the adiabatic theorem for periodically driven systems.
Let us consider a periodically pulsed driven system (with a period $T$)
described by the ``kicked'' Hamiltonian:
\begin{equation}
  \label{eq:kickedHamiltonian}
  \hat{H}(t) = \hat{H}_0 + \lambda\hat{V}\sum_{n\in\Integer}\delta(t - nT),
\end{equation}
where $\hat{H}_0$ and $\hat{V}$ describe
the ``free'' motion and the pulsed perturbation, respectively,
and $\lambda$ is the strength of the perturbation.
In the following, we focus on the stroboscopic description of the 
state vector $\ket{\psi_n}$ at $t = nT-0$. The time evolution of 
$\ket{\psi_n}$ is described by the quantum map
$\ket{\psi_{n+1}}=\hat{U}_{\lambda}\ket{\psi_{n}}$, where
\begin{equation}
  \label{eq:defUlambda}
  \hat{U}_{\lambda} = e^{-i \hat{H}_0 T/\hbar} e^{-i\lambda\hat{V}/\hbar}
\end{equation}
is a Floquet operator~\cite{QuantumMap}.
In the following, we set $\hbar = 1$.
In order to show a simple example of the parametric motions of 
eigenvalues and eigenvectors, we assume that 
the spectrum of $\hat{U}_{\lambda}$ contains only
discrete components and has no degeneracy.
At the same time, in order to avoid subtle issues that are 
brought from the infinite dimensionality 
of the Hilbert space $\Hilbert$~\cite{endnote:infinite},
we assume that $N\equiv\dim\Hilbert$ is finite.
This assumption does no harm to the descriptions of many systems
where an appropriate introduction of the truncation of the Hilbert space 
is feasible.

Since $\hat{U}_{\lambda}$ (\ref{eq:defUlambda}) is unitary,
its eigenvalues $\{z_n(\lambda)\}_{n=0}^{N-1}$
are complex and on the unit circle, i.e., $|z_n(\lambda)|=1$. 
The phase of $z_n(\lambda)$ indicates
the increment of the dynamical phase during the unit time evolution
whose initial state is the corresponding eigenstate $\ket{\xi_n(\lambda)}$.
The time-average of the dynamical phase 
determines a quasienergy 
$E_n(\lambda) = 
- T^{-1} {\rm Im}[\ln z_n(\lambda)]
$
(or, $z_n(\lambda) = e^{-i E_n(\lambda) T}$).
Note that the value of quasienergy has an ambiguity because of the period 
$2\pi/T$ in the quasienergy space.
We remark that the eigenvalue equation
\begin{equation}
  \label{eq:eigenvalueEq}
  \hat{U}_{\lambda} \ket{\xi_n(\lambda)}
  = z_n(\lambda) \ket{\xi_n(\lambda)}
\end{equation}
determines the eigenvalue and the eigenvector only pointwise 
in $\lambda$.
By assuming the continuity about $\lambda$, we obtain the derivatives of 
$E_n(\lambda)$ and $\ket{\xi_n(\lambda)}$~\cite{Nakamura:PRA-35-5294}:
\begin{eqnarray}
  \label{eq:leveldynamics}
  \pdfrac{}{\lambda} E_n(\lambda) &=&
  \frac{1}{T} \braOket{\xi_n(\lambda)}{\hat{V}}{\xi_n(\lambda)},
  \\
  \label{eq:vectordynamics}
  \pdfrac{}{\lambda} \ket{\xi_n(\lambda)} &=&
  -i A_n(\lambda)\ket{\xi_n(\lambda)}
  \nonumber \\ && 
  {}+ i\sum_{m\ne n}
  \frac{z_m(\lambda) \braOket{\xi_m(\lambda)}{\hat{V}}{\xi_n(\lambda)}}%
  {z_m(\lambda) - z_n(\lambda)}
  \ket{\xi_m(\lambda)},
  \nonumber\\
\end{eqnarray}
where $\ket{\xi_n(\lambda)}$ is assumed to be normalized and
$A_n(\lambda) \equiv 
i \bra{\xi_n(\lambda)}\partial\ket{\xi_n(\lambda)}/\partial \lambda$
is a geometric gauge potential~\cite{Mead:JCP-70-2284}.
These derivatives compose a set of equations of motion for a 
virtual time $\lambda$~\cite{endnote:leveldynamics}.
With a given ``initial condition'' of
$\{E_n(\lambda), \ket{\xi_n(\lambda)}\}_n$, at $\lambda = \lambda_0$,
we may integrate the equations of motion~(\ref{eq:leveldynamics}) and
(\ref{eq:vectordynamics}).
Note that, under the presence of 
anholonomy~\cite{GeometricPhaseReview},
the single-valuedness of the solution 
$\{E_n(\lambda), \ket{\xi_n(\lambda)}\}_n$ generally holds only locally 
in the parameter space of $\lambda$.

The adiabatic theorem~\cite{Born:ZP-51-165}
for periodically driven systems~\cite{Holthaus:PRL-69-1596}
provides 
a physical significance of the geometry (i.e., $\lambda$-dependence) of 
$\ket{\xi_n(\lambda)}$. 
Note that the parameter $\lambda$, which is supposed to be slowly changed,
is the strength of the perturbation that is applied periodically:
$\lambda$ will be changed from $\lambdai$ to $\lambdaf$, during
the $\Nt$ steps, where the corresponding time interval is $T\Nt$. 
Let $\lambda_j$ be the value of $\lambda$ at the $j$-th step 
($0 \le j \le \Nt$). 
In particular $\lambda_0 = \lambdai$ and
$\lambda_{\Nt} = \lambdaf$. 
The slowness of the change of the parameter is
expressed by the condition 
$\lambda_{j+1} - \lambda_j = \mathcal{O}(\Nt^{-1})$ as $\Nt\to\infty$. 
We start with an initial condition that, at $\lambda=\lambdai$, 
the system is in an
eigenstate $\ket{\xi_n(\lambdai)}$ of $\hat{U}_{\lambdai}$. 
The final state $\ket{\Psi_{\rm f}}$ is
\begin{equation}
  \ket{\Psi_{\rm f}} \equiv
  \Torder\left[\prod_{j=1}^{\Nt} \hat{U}_{\lambda_j}\right] 
  \ket{\xi_n(\lambdai)},
\end{equation}
where $\Torder$ represents a time-orderd (or, equivalently, path-orderd) 
product.
According to the adiabatic theorem,
the final state will converge to
$\ket{\xi_n(\lambdaf)}$ as $\Nt\to\infty$, except its phase.
In the following, we will evaluate the phase of the final state.
From the equation of motion~(\ref{eq:vectordynamics}), we have
\begin{equation}
  \begin{split}
    &\hat{U}_{\lambda} \ket{\xi_n(\lambda - \delta)}
    \\&=
    \exp\left\{-i E_n(\lambda)T + i A_n(\lambda)\delta\right\}
    \ket{\xi_n(\lambda)} 
    \\ &\quad
    - \sum_{m\ne n} 
    \frac{iz_m(\lambda)^2\bra{\xi_m(\lambda)}\hat{V}\ket{\xi_n(\lambda)}\delta}{z_m(\lambda) - z_n(\lambda)}\ket{\xi_m(\lambda)}
    +\mathcal{O}(\delta^2)
  \end{split}
\end{equation}
as $\delta\to0$. 
According to the adiabatic theorem~\cite{Holthaus:PRL-69-1596}, 
we need only the first term above for the evaluation of the phase.
Hence we have
\begin{equation}
  \ket{\Psi_{\rm f}}
  \simeq
  \exp \left\{
    - i \sum_{j=1}^{\Nt} E_n(\lambda_j) T
    + i \int_{\lambdai}^{\lambdaf} A_n(\lambda) d\lambda
  \right\}
  \ket{\xi_n(\lambdaf)},
\end{equation}
as $\Nt\to\infty$.
The first and the second terms in the phase factor correspond to the dynamical 
and geometric phases, respectively~\cite{Berry:PRSLA-430-405}.

\section{Quantum map under a rank-$1$ perturbation}
\label{sec:rank1}

In order to demonstrate the simplest example of Cheon's 
anholonomies in quantum maps, we employ a rank-$1$ perturbation
$\hat{V}=\ket{v}\bra{v}$ in Eq.~(\ref{eq:defUlambda})
with a normalized vector $\ket{v}$~\cite{Combescure:JSP-59-679}.
Since $\hat{V}$ satisfies $\hat{V}^2=\hat{V}$, 
the quantum map~(\ref{eq:defUlambda}) has a periodicity about $\lambda$. 
This is shown by an expansion of $\hat{U}_{\lambda}$ in $\hat{V}$,
\begin{equation}
  \label{eq:expandUlambda}
  \hat{U}_{\lambda} = 
  \hat{U}_{0} \left\{1 - (1 - e^{-i\lambda}) \hat{V}\right\},
\end{equation}
which has  a $2\pi$ periodicity in $\lambda$.
Hence the parameter space of $\lambda$ is identified with a circle $S^1$.
We will discuss the parametric motion of quasienergies and 
eigenvectors of $\hat{U}_{\lambda}$, along the changes of $\lambda$ on $S^1$.

Two kinds of ``trivial'' eigenvectors of $\hat{U}_{\lambda}$ are shown
in order to simplify the later analysis on Cheon's anholonomies.
For the first kind, we suppose that an eigenvector 
$\ket{\xi}$ of $\hat{U}_{\lambda_0}$ 
is orthogonal to $\ket{v}$ and the corresponding eigenvalue is $z_0$. 
Then, this implies that $\ket{\xi}$ is
also an eigenvector of $\hat{U}_{\lambda}$ for all $\lambda$
and the corresponding eigenvalue $z_0$ does not depend on $\lambda$.
In fact, we have
\begin{equation}
  \label{eq:orthoTriviall}
  \hat{U}_{\lambda}\ket{\xi}
  = \hat{U}_{\lambda_0}e^{-i(\lambda - \lambda_0)\hat{V}/\hbar}\ket{\xi}
  = \hat{U}_{\lambda_0}\ket{\xi}
  = z_0 \ket{\xi},
\end{equation}
where we used $\hat{V}\ket{\xi} =\ket{v}\bracket{v}{\xi}=0$ 
and $\hat{U}_{\lambda_0}\ket{\xi} = z_0 \ket{\xi}$.
In Appendix~\ref{sec:normalForm},
we will show that such trivial eigenvectors~(\ref{eq:orthoTriviall})
are created by a spectral degeneracy of $\hat{U}_{\lambda}$.
For the second kind, we suppose that $\ket{v}$ is an eigenvector 
of $\hat{U}_{\lambda_0}$ 
and the corresponding eigenvalue is $z_0$. 
If this is the case, all the eigenvectors of 
$\hat{U}_{\lambda_0}$, except $\ket{v}$, are orthogonal to $\ket{v}$, and 
accordingly become trivial eigenvectors of the first kind mentioned above.
Furthermore, $\ket{v}$ is also a trivial one in the sense that
$\ket{v}$ is an eigenvector of $\hat{U}_{\lambda}$ for all $\lambda$.
This is because
\begin{eqnarray}
  \label{eq:paraTriviall}
  \hat{U}_{\lambda}\ket{v}
  &=&
  \hat{U}_{\lambda_0}e^{-i(\lambda - \lambda_0)\hat{V}/\hbar}\ket{v}
  = \hat{U}_{\lambda_0} e^{-i(\lambda - \lambda_0)/\hbar}\ket{v}
  \nonumber \\
  &=&
  z_0 e^{-i(\lambda - \lambda_0)/\hbar} \ket{v},
\end{eqnarray}
where the corresponding eigenvalue 
$z_0 e^{-i(\lambda - \lambda_0)/\hbar}$ depends on $\lambda$.
The analysis of the two kinds of trivial eigenvectors 
are completed.

In the following, we assume the absence of these trivial eigenvectors 
since they
are irrelevant to the later argument to look for Cheon's 
anholonomies. A systematic procedure to reduce a Hilbert space
by excluding these trivial eigenvectors of $\hat{U}_{\lambda}$
is explained in Appendix~\ref{sec:normalForm}.
On the reduced Hilbert space $\Hilbert$, it is assured that the spectrum of
$\hat{U}_{\lambda}$ has no degeneracies for all $\lambda$.
In terms of $\hat{U}_0$ and $\ket{v}$, 
this assumption turns out to be equivalent to the following two conditions.
(i) The spectrum of $\hat{U}_0$ is nondegenerate.
Note that we have already introduced another assumption that
the spectrum of $\hat{U}_0$ contains only discrete and 
a finite number of components to assure the smoothness of the parametric 
dependence of eigenvalues and eigenvectors on $\lambda$
in Section~\ref{sec:adiabaticTransport}.
(ii) $\ket{v}$ is not orthogonal to any eigenvector of $\hat{U}_0$, 
otherwise the reduction is not complete.
Note that (ii) implies that $\ket{v}$ is not any eigenvector of $\hat{U}_0$.
Thus the conditions (i) and (ii) guarantee, for all $\lambda$,
\begin{equation}
  \label{eq:v_xi}
  0 \lvertneqq |\bracket{v}{\xi}| \lvertneqq 1
  \quad\text{for any eigenvector $\ket{\xi}$ of $\hat{U}_{\lambda}$},
\end{equation}
where either the lower or the upper bound of the equalities would hold 
if any trivial eigenvector remains.

The conditions (i) and (ii) are further paraphrased with the help of
the notion {\em cyclicity}~\cite{RSICyclicity}, when we restrict 
ourselves to
the dimensionality of the Hilbert space $\Hilbert$ being finite. 
If $\hat{U}_0$ and $\ket{v}$ satisfy
$\Hilbert = \overline{{\rm span} \{(\hat{U}_0)^m \ket{v}\}_{m=0}^{\infty}}$%
~\cite{note:infiniteSpan},
$\ket{v}$ is called a cyclic vector for $\hat{U}_0$~\cite{RSICyclicity}.
It is shown in Appendix~\ref{sec:cyclicity}, the conditions (i) and (ii) are 
equivalent to
(i') $\hat{U}_0$ has a cyclic vector
and (ii') $\ket{v}$ is a cyclic vector for $\hat{U}_0$,
respectively.
We will discuss the case that these assumptions are broken
in Section~\ref{sec:examples}.

\section{Cheon's anholonomies in quantum maps}
\label{sec:existance}

What happens to the quasienergies and the eigenvectors
when we adiabatically increase $\lambda$ by $2\pi$, the period of
the Floquet operator~(\ref{eq:expandUlambda}), and its spectrum, 
starting from $\lambda=\lambda_0$? 
The argument above suggests that we may have a conventional
(Abelian) phase anholonomy that appears only in the phase of 
the eigenvectors. 
However, the following argument will elucidate that we meet 
Cheon's anholonomies in quasienergies as well as in eigenspaces.

First, we examine quasienergies $E_n(\lambda)$ ($0\le n <N = \dim\Hilbert$).
Note that we have $N$ cases to choose ``the ground quasienergy''
due to the periodicity in the quasienergy space.
Once we choose a ground state, whose quantum number is assigned to $0$,
the quantum number $n$ ($<N$) is assigned so 
that $E_n(\lambda)$ increases as $n$ increases.
More precisely, in order to remove ambiguities due to the periodicity in
the quasienergies, we choose the branch of the quasienergies, 
at $\lambda=\lambda_0$, 
as $E_0(\lambda_0) < E_1(\lambda_0) < \cdots < E_{N-1}(\lambda_0)
< E_0(\lambda_0)+2\pi T^{-1}$ holds, 
where $E_{0}(\lambda_0)$ and $E_0(\lambda_0)+2\pi T^{-1}$ correspond
to the same eigenvalue $z_0(\lambda_0) = e^{-i E_{0}(\lambda_0) T}$.
For brevity, we identify a quantum number $n$ with $n+N$.

To examine how much the ground quasienergy $E_0(\lambda_0)$ increases 
during a cycle of $\lambda$, we evaluate
\begin{equation}
  \label{eq:DEnIE}
  \Delta E_n \equiv 
  \int_{\lambda_0}^{\lambda_0+2\pi}
  \pdfrac{E_n(\lambda)}{\lambda} d\lambda.
\end{equation}
Note that $\Delta E_n$ is ``quantized''
due to the periodicity of the spectrum, e.g., we have
\begin{equation}
  \label{eq:quantizedDE0}
  \Delta E_0 =  E_{\nu}(\lambda_0) - E_0(\lambda_0) \mod 2\pi T^{-1}
  \quad\text{for some $\nu$},
\end{equation}
because $E_0(\lambda)$ should arrive at $E_{\nu}(\lambda_0)$ 
for some $\nu$ as $\lambda\nearrow \lambda_0+2\pi$.
To determine which $\nu$ is possible or not, 
we evaluate the integral expression~(\ref{eq:DEnIE}) of $\Delta E_n$ with 
$\partial {E_n(\lambda)}/\partial\lambda
= T^{-1} \braOket{\xi_n(\lambda)}{\hat{V}}{\xi_n(\lambda)}$~%
\cite{Nakamura:PRA-35-5294}.
Since $\hat{V}$ satisfies $\hat{V}^2=\hat{V}$, 
the eigenvalues of $\hat{V}$ are only $0$ and $1$.
Accordingly we have 
$0 \le \partial {E_n(\lambda)}/\partial\lambda \le T^{-1}$.
However, the equalities for the minimum and the maximum cannot hold
because of $ \partial{E_n(\lambda)}/\partial\lambda
= T^{-1} |\bracket{v}{\xi_n(\lambda)}|^2$
and 
$0 < |\bracket{v}{\xi_n(\lambda)}| < 1$ (see  Eq.~(\ref{eq:v_xi})).
Hence we have 
$
  0 < \partial {E_n(\lambda)}/\partial\lambda < T^{-1}
$
and accordingly
\begin{equation}
  \label{eq:DeltaEnRestriction}
  0 < \Delta E_n < 2\pi T^{-1}.
\end{equation}
This imposes a restriction $0 < \nu < N$ in Eq.~(\ref{eq:quantizedDE0}).
In particular, neither $\nu = 0$ nor $N$ is possible.
Namely, the quasienergy $E_0(\lambda)$ arrives at
$E_{\nu}(\lambda_0)$ ($0 < \nu < N$), instead of $E_0(\lambda_0)$, 
as $\lambda\nearrow \lambda_0+2\pi$.
This is nothing but a manifestation of Cheon's anholonomy in quasienergy.

If the system is two-level (i.e., $N = 2$), 
the above argument immediately implies $\nu = 1$.
Hence it is straightforward to show that
\begin{equation}
  \label{eq:quasienergyAnholonomyRank1}
  E_n(\lambda_0 + 2\pi-0) = E_{n+1}(\lambda_0) \mod 2\pi T^{-1}
\end{equation}
holds for all $0 \le n < N$.

Equation~(\ref{eq:quasienergyAnholonomyRank1}) remains true for $N>2$.
Its justification requires one to examine a sum rule on 
$\set{\Delta E_n}_{n=0}^{N-1}$:
\begin{equation}
  \label{eq:sumRuleRank1}
  \sum_{n=0}^{N-1}\Delta E_n 
  = \int_{\lambda_0}^{\lambda_0 + 2\pi} \frac{1}{T} ({\rm Tr} \hat{V})d\lambda
  =\frac{2\pi}{T},
\end{equation}
where we used ${\rm Tr} \hat{V} = 1$ for $\hat{V}=\ket{v}\bra{v}$
with normalized $\ket{v}$.
The sum rule~(\ref{eq:sumRuleRank1}) implies that
Eq.~(\ref{eq:quasienergyAnholonomyRank1}) holds for all $0 \le n < N$,
and vice versa,
where the sum $\sum_{n=0}^{N-1}\Delta E_n$ in Eq.~(\ref{eq:sumRuleRank1})
takes its possible minimal value $2\pi T^{-1}$.
Actually, if we assume that
Eq.~(\ref{eq:quasienergyAnholonomyRank1}) is broken for some $n$,
e.g.,
$E_{n}(\lambda_0 + 2\pi-0) = E_{n+\nu}(\lambda_0) \mod 2\pi/T$
with $1 < \nu < N$, 
this contradicts with the sum rule~(\ref{eq:sumRuleRank1}).
Thus the quasienergy anholonomy~(\ref{eq:quasienergyAnholonomyRank1})
for $N$-level quantum maps under 
the rank-1 perturbation is revealed completely.

The quasienergy anholonomy~(\ref{eq:quasienergyAnholonomyRank1}) induces 
an eigenspace anholonomy, which is expressed by projectors:
\begin{equation}
\ket{\xi_n(\lambda_0 + 2\pi-0)}\bra{\xi_n(\lambda_0 + 2\pi-0)}
= \ket{\xi_{n+1}(\lambda_0)}\bra{\xi_{n+1}(\lambda_0)}.
\end{equation}
Note that $\ket{\xi_{n}(\lambda_0)}$ and $\ket{\xi_{n+1}(\lambda_0)}$ 
are orthogonal, since the corresponding eigenvalues are different.

Finally, we show an anholonomy in a state vector as a result of
the adiabatic increment of $\lambda$ by the period $2\pi$
from $\lambda=\lambda_0$.
When the initial state is prepared to be
an eigenstate $\ket{\xi_n(\lambda_0)}$, the corresponding final state is 
\begin{equation}
  \exp \left\{- i \sum_{j=1}^{\Nt} E_n(\lambda_j) T
    + i \int_{\lambda_0}^{\lambda_0+2\pi} A_n(\lambda) d\lambda
  \right\}
  \ket{\xi_{n+1}(\lambda_0)}.
\end{equation}
If we keep the adiabatic increment of $\lambda$, the state vector
will become parallel with the eigenvector 
$\ket{\xi_{n+\nu}(\lambda_0)}$ of $\hat{U}_{\lambda_0}$
after the completion of the $\nu$-th iteration of the periodic increment
and return to the initial eigenstate at the end of the $N$-th iteration.

\section{Examples}
\label{sec:examples}
\begin{figure}
  \includegraphics[width=8.6cm]{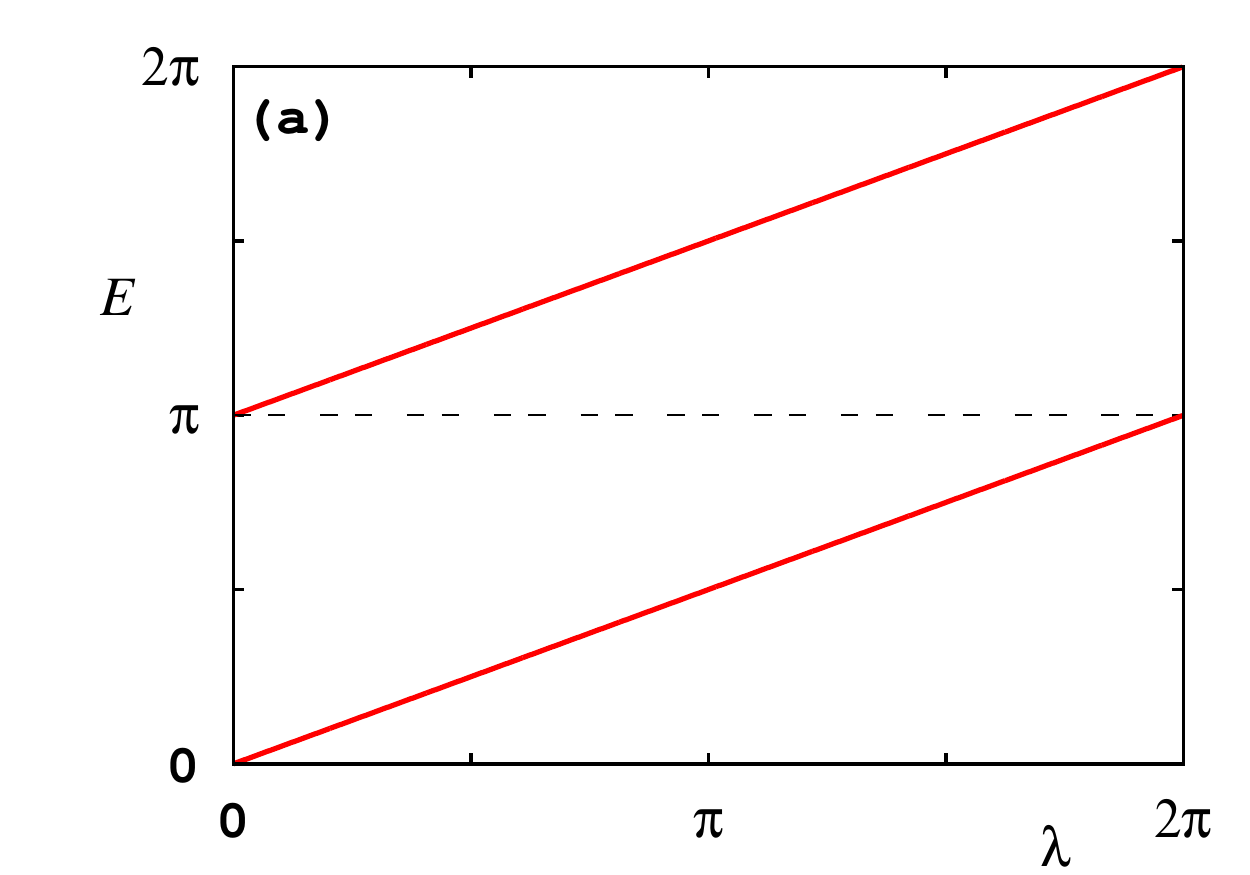}
  \hfill
  \includegraphics[width=8.6cm]{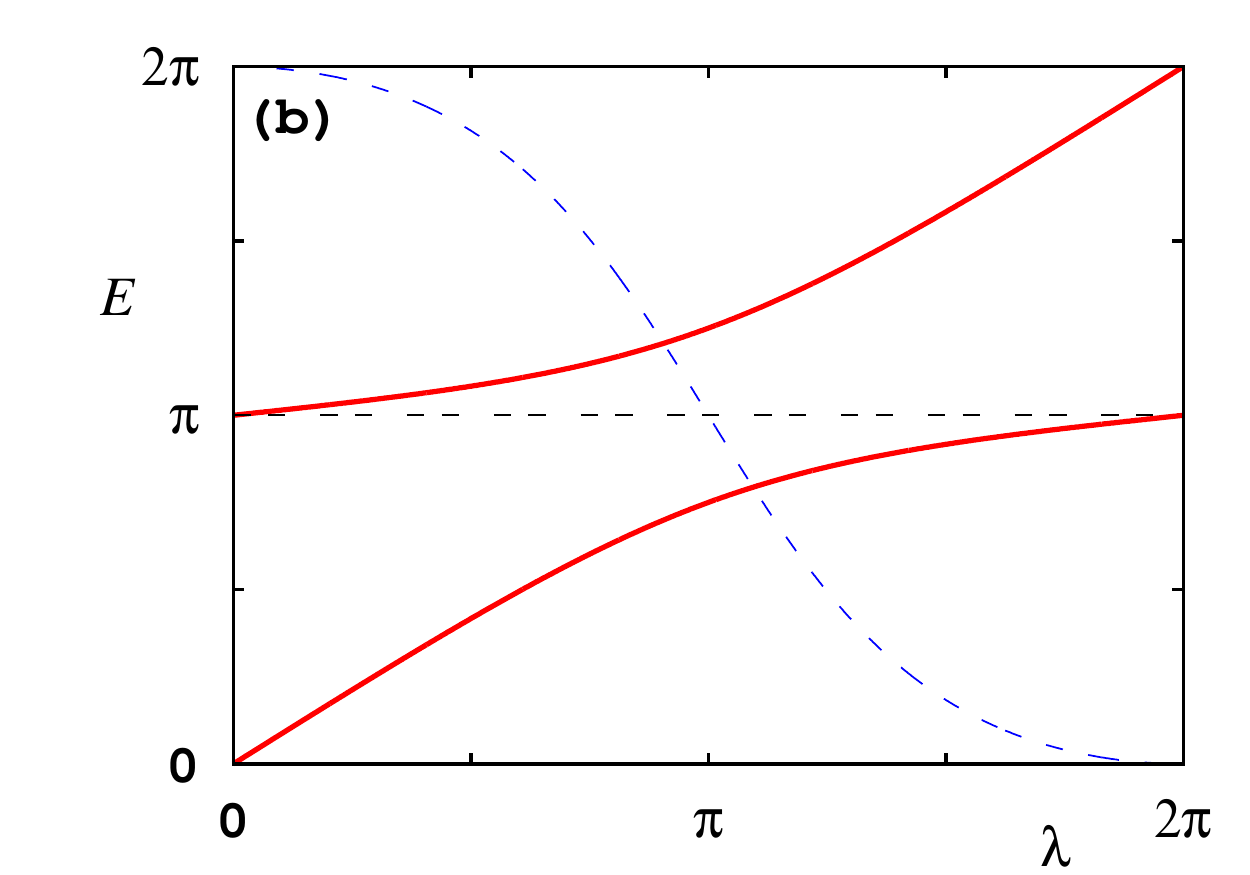}
  \caption{\label{fig:twolevel} (Color online)
    Parametric motions of quasienergies (bold lines) of two-level systems.
    We choose the model whose ``ground'' quasienergy is zero at $\lambda = 0$.
    The other quasienergy $\delta=\pi$ at $\lambda = 0$ is indicated 
    by the broken, horizontal lines.
    (a) The example examined in the main text, Eq.~(\ref{eq:sampleE})
    ($\ket{v}=(\ket{\up}-i\ket{\down})/\sqrt{2}$).
    The quasienergies draw two parallel lines, which have no
    avoided crossing.
    (b) A generic example 
    ($\ket{v} = \cos(\pi/8)\ket{\up} + \sin(\pi/8)\ket{\down}$). 
    There is a single avoided crossing. The broken curve represents
    $|\bracket{\up}{\xi_0(\lambda)}|^2$, which depicts
    that $\ket{\xi_0(\lambda)}$ 
    becomes orthogonal to $\ket{\xi_0(0)}$ in
    the limit $\lambda \uparrow 2\pi$.
  }
\end{figure}

The simplest example of Cheon's anholonomies occurs in a two-level system.
The Floquet operator of the unperturbed system is
\begin{equation}
  \hat{U}_0 \equiv \ketbra{\up}{\up} + \ket{\down}e^{-i\delta}\bra{\down},
\end{equation}
where $\delta\in(0,2\pi)$ and $T=1$ are assumed.
We employ $\ket{v}=(\ket{\up}-i\ket{\down})/\sqrt{2}$, which satisfies
the conditions (i) and (ii) mentioned in Section~\ref{sec:rank1}.
Although the constant term in 
$\hat{V}=\ket{v}\bra{v} = \frac{1}{2}(1-\hat{\sigma}_y)$
seems to be
irrelevant, this term arises naturally in projection operators
in the two-level system and is required to ensure that 
the perturbed Floquet operator 
$\hat{U}_{\lambda}=\hat{U}_{0} e^{-i\lambda\hat{V}}$
has the $2\pi$ periodicity about $\lambda$.
In order to show an analytic form of quasienergies and eigenvectors,
we employ $\delta=\pi$ (i.e., $\hat{U}_0 = \hat{\sigma}_z$).
The eigenvalues of $\hat{U}_{\lambda}$ are
$z_{0}(\lambda) = e^{-i\lambda/2}$ and
$z_{1}(\lambda) = -e^{-i\lambda/2}$.
The period of each eigenvalue about $\lambda$ is $4\pi$, 
though the period of the spectrum $\{z_{0}(\lambda), z_{1}(\lambda)\}$ is
$2\pi$.
The corresponding quasienergies are 
\begin{equation}
  \label{eq:sampleE}
  E_{0}(\lambda)= \frac{1}{2}\lambda\mod 2\pi
  \quad\text{and}\quad
  E_{1}(\lambda)= \pi + \frac{1}{2}\lambda\mod 2\pi.
\end{equation}
Now we demonstrate the anholonomy in quasienergy (see Fig.~\ref{fig:twolevel}):
At $\lambda=0$, we start from the quasienergy $E_{0}(0) = 0$ of the 0-th 
eigenstate. 
The increment of $\lambda$ increases
$E_0(\lambda)$ because of the fact $dE_0(\lambda)/d\lambda = \frac{1}{2} > 0$.
At $\lambda=2\pi$, $E_{0}(\lambda)$ arrives at 
$\pi$, which agrees with the quasienergy $E_1(0) = \pi$
of the first eigenstate at $\lambda=0$.
Next, we examine the eigenvectors
\begin{align}
  \label{eq:twoleveleigenvecs}
  \ket{\xi_0(\lambda)} = 
  \begin{bmatrix}\cos(\lambda/4)\\ \sin(\lambda/4)\end{bmatrix},
  \quad
  \ket{\xi_1(\lambda)} = 
  \begin{bmatrix}-\sin(\lambda/4)\\ \cos(\lambda/4)\end{bmatrix}.
\end{align}
The corresponding geometric gauge potentials $A_{n}(\lambda)$ ($n=0,1$)
happen to vanish in the present case. Hence it is easy to 
find the geometric phases from 
the parametric dependence of the eigenvectors~(\ref{eq:twoleveleigenvecs}).
The excursion of the eigenvectors by increasing the parameter $\lambda$
is the following:
\begin{align}
  \ket{\xi_{0}(0)} &= \ket\up,&\quad
  \ket{\xi_{1}(0)} &= \ket\down,
  \nonumber \\
  \ket{\xi_{0}(2\pi)} &= \ket{\xi_{1}(0)},&\quad
  \ket{\xi_{1}(2\pi)} &= -\ket{\xi_{0}(0)},
  \\
  \ket{\xi_{0}(4\pi)} &= - \ket{\xi_{0}(0)},&\quad
  \ket{\xi_{1}(4\pi)} &= -\ket{\xi_{1}(0)},
  \nonumber
\end{align}
where nontrivial geometric phases appear after the completion
of the $4\pi$ increment of $\lambda$.

We suggest a possible implementation of
the example above in a charged particle with a spin-$1/2$.
Assume that the particle is localized to some place so that we may
ignore the motion of the particle.
The unperturbed system is the spin under a static magnetic 
field. The perturbation $\hat{V}$ is composed of two ingredients: 
One is a periodically pulsed magnetic field, whose direction needs to be 
different from that of the unperturbed magnetic field. 
The other is 
a periodically pulsed electric field, which provides 
``the constant part'' of $\hat{V}$. In order to prepare $\hat{V}$, 
we need to adjust the ratio of the strength and the period of 
the two perturbation fields.

\begin{figure}
  \includegraphics[width=8.6cm]{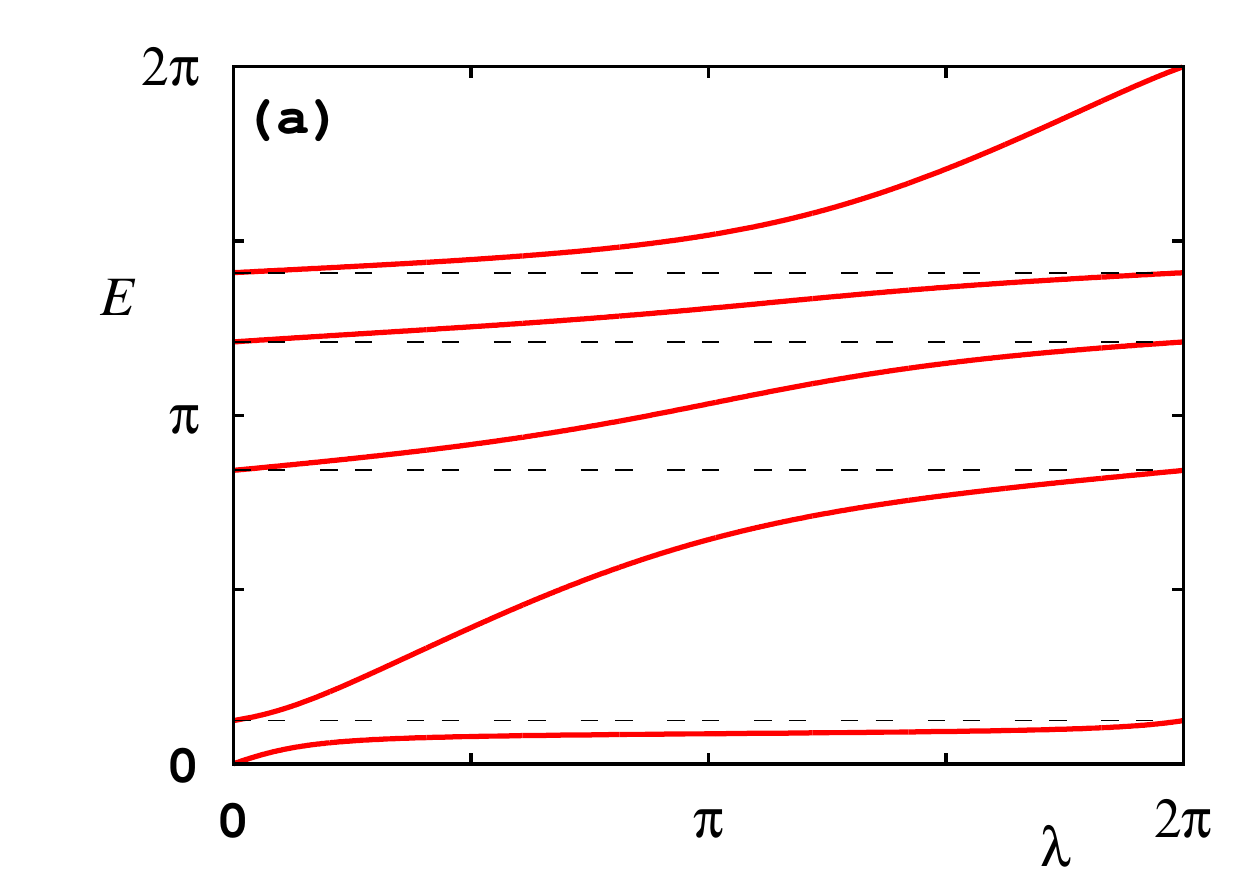}
  \hfill
  \includegraphics[width=8.6cm]{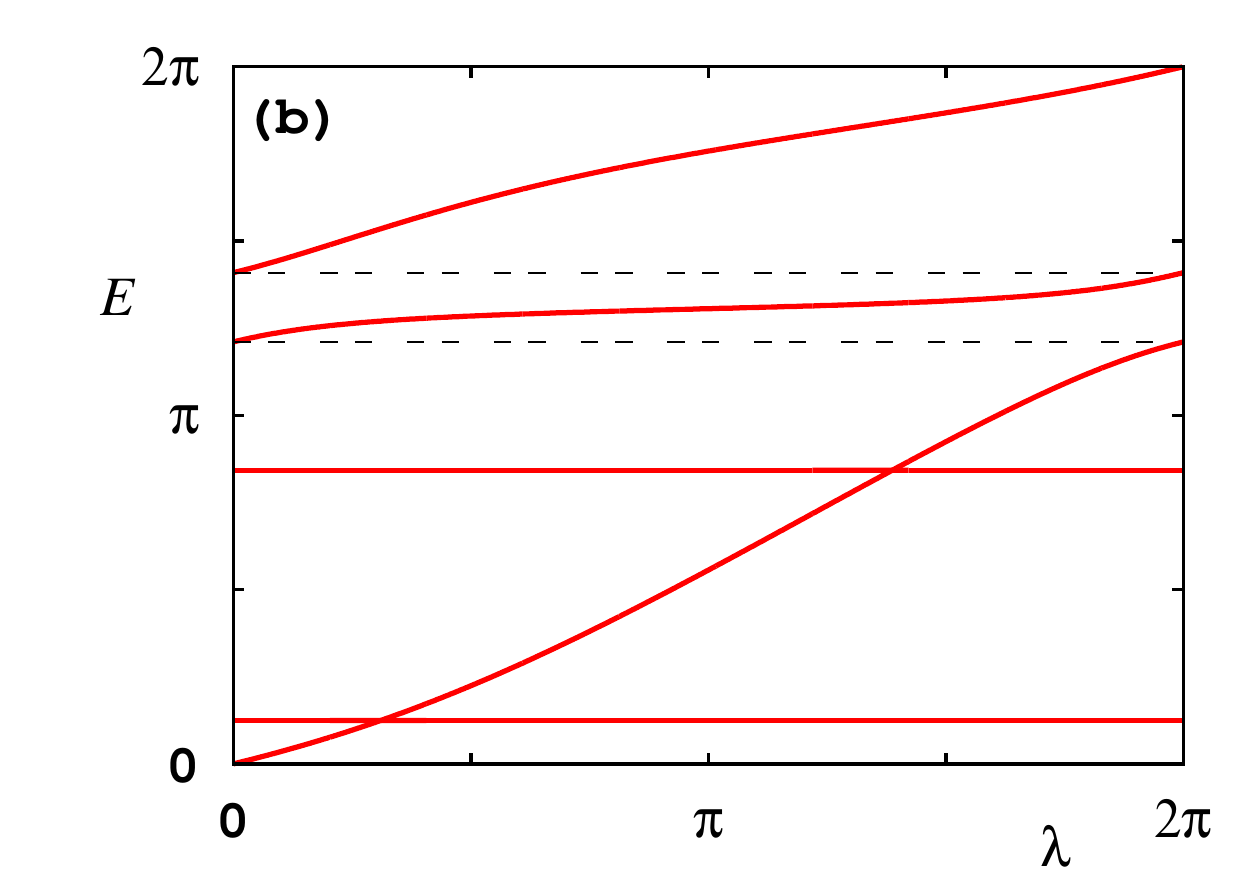}
  \caption{\label{fig:multilevel} (Color online)
    Parametric motions of quasienergies (bold lines) in systems 
    with multiple levels ($\dim\Hilbert = 5$).
    The unperturbed Floquet operator $\hat{U}_0$,
    whose ``ground'' quasienergy is adjusted to zero at $\lambda = 0$,
    is randomly chosen.
    The quasienergies at $\lambda = 0$ are indicated by the broken, 
    horizontal lines.
    (a) A random choice of $\ket{v}$, which satisfies cyclicity for
    $\hat{U}_0$.
    All quasienergies exhibit anholonomy.
    (b) An example for broken cyclicity in $\ket{v}$:
    Only three components of $\ket{v}$, in the representation
    that diagonalizes $\hat{U}_0$, take nonzero values.
    The resultant parametric changes and anholonomy occur only in
    the subspace ${\rm span} \set{\hat{U}_0^m\ket{v}}_{m}$,
    whose dimensionality is three.
    The other two quasienergies draw horizontal lines,
    since they correspond to the trivial eigenvectors mentioned
    in Section~\ref{sec:rank1},
    and are not affected by the perturbation.
  }
\end{figure}

Finally, we show another example that involves multiple levels
in Fig.~\ref{fig:multilevel} (a), where all quasienergies are involved 
in the anholonomy. 
This is due to the cyclicity of $\ket{v}$.
In Fig.~\ref{fig:multilevel}~(b), we also show an
example that breaks the cyclicity of $\ket{v}$.
This suggests that we may control the anholonomy to the limited number of
states by an appropriate choice of $\ket{v}$. 

\section{Geometry and abundance of quasienergy anholonomy}
\label{sec:abundance}

\begin{figure}
  \includegraphics[width=4.22cm]{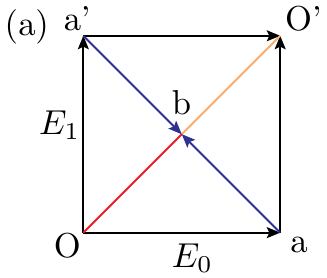}
  \ 
  \includegraphics[width=4.22cm]{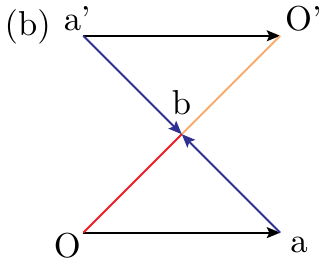}
  \\[\baselineskip]

  \includegraphics[width=4.22cm]{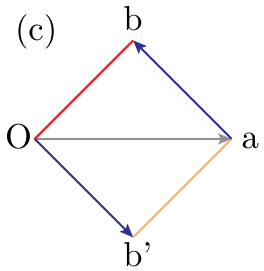}
  \ 
  \includegraphics[width=4.22cm]{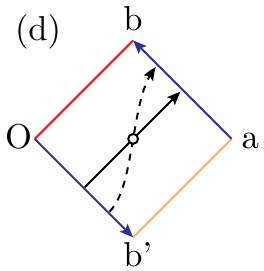}

  \caption{\label{fig:geometry} (Color online)
    A parametrization with the quasienergy-spectrum of quantum systems whose 
    unit time evolutions are described by two-dimensional Floquet operators
    is explained.
    (a) An $E_0-E_1$ plane, where a and a' are at 
    $(0, 2\pi/T)$ and $(2\pi/T, 0)$, respectively.
    Because of the periodicity of quasienergies, pairs of lines
    Oa and a'O', and, Oa' and aO' are identified. 
    On the diagonal line $E_0 = E_1$, two quasienergies are degenerate.
    In the subsequent figures, $(E_0, E_1)$ and $(E_1, E_0)$ are identified.
    (b) $\triangle$Oba' and $\triangle$O'ba are identified with
    $\triangle$Oba and $\triangle$O'ba', respectively, and so are removed.
    (c) The quotient space  $T^2/\Integer_2$.
    Since Oa and a'O' in (b) are identical, they are arranged to make
    a square. Furthermore, if identical lines Ob' and ab are put
    together, a M\"obius strip with edges Ob and ab', which are
    degenerate lines, is obtained
    (see Ref.~\cite{Tsutsui:JMP-42-5687}).
    (d) Parametric motion of spectrum on $T^2/\Integer_2$.
    Bold and dashed lines, which are topologically equivalent
    on the M\"obius strip, correspond to 
    Figs.~\ref{fig:twolevel} (a) and (b), respectively, where
    $T=1$ is assumed. At $\lambda=0$,
    they start from the open circle at $(0, \pi)$. As $\lambda$ increases,
    they move toward ab. At $\lambda=\pi$, they arrive on a point on a line
    ab, which is identical with $Ob'$. Then they return to $(0, \pi)$.
    Such a winding along the M\"obius strip induces the quasienergy 
    anholonomy. 
  }
\end{figure}

We remark on the geometry of the quasienergy anholonomy to discuss its
stability and abundance.
Concerning Cheon's eigenenergy anholonomy in a family of systems with 
generalized pointlike potentials, 
Tsutsui, F\"ulop, and Cheon examined the geometry of the anholonomy
using the fact that the parameter space of the family is 
$U(2)$~\cite{Cheon:AP-294-1,Tsutsui:JMP-42-5687}.
When the dimension of the Hilbert space is two,
Tsutsui {\it et al.}'s argument is immediately applicable to the 
quasienergy anholonomy in the systems whose unit time evolution is 
described by a Floquet operator, which is a $2\times2$ unitary
matrix.
We employ a parametrization of such systems by
their quasienergy-spectrum $\{(E_0, E_1)\}$~(Fig.~\ref{fig:geometry} (a)), 
whose element $(E_0, E_1)$ is identified with $(E_1, E_0)$. 
The quotient quasienergy-spectrum space is accordingly
an orbifold $T^2/\Integer_2$
which has two topologically inequivalent and nontrivial 
cycles (see  Fig.~\ref{fig:geometry} (c) and Ref.~\cite{Tsutsui:JMP-42-5687}). 
One cycle traverses the degeneracy line $E_0 = E_1$. 
The other cycle concerns the ``increment'' (or ``decrement'') of 
the quantum number.
More precisely,
the winding number along the latter cycle determines the increment of 
the quantum number (Fig.~\ref{fig:geometry} (d)).
When the dimension of Hilbert space is  larger than 2,
similar geometrical argument will be possible.
The geometrical nature implies that the quasienergy anholonomy is
stable against perturbations that preserves the topology of the cycle.
Hence we may expect that the same anholonomy appears in
nonautonomous systems whose unit time evolution is described by 
a Floquet operator, e.g., periodically kicked systems and periodically driven 
systems.

The stability of Cheon's anholonomies against perturbations is also 
expected from the fact 
that the parametric dependence of the quasienergies has no crossings 
(see Figs.~\ref{fig:twolevel} and \ref{fig:multilevel}~(a)).
To achieve the stability in practice, the gap of narrowly avoided 
crossings needs to be enlarged. This is possible with a
suitable adjustment of $\ket{v}$ 
(e.g., see Figs.~\ref{fig:twolevel} (a) and (b)).
We also remark that the presence of the trivial eigenvectors, 
which are introduced in Section~\ref{sec:rank1},
induces the crossings of quasienergies, as can be seen in 
Fig.~\ref{fig:multilevel}~(b) and in Appendix~\ref{sec:normalForm}.
Hence quasienergy anholonomies coexisting with trivial eigenvectors 
are fragile against perturbations in general.

\section{Application: anholonomic quantum state manipulation}
\label{sec:manipulation}

As an application of the quasienergy anholonomy, a design principle of 
systems that achieve manipulations of quantum states with adiabatic passages 
is proposed.
Before describing our argument, we mention that the conventional 
works on the application of adiabatic passages to the manipulation of
quantum states are the textbook results~\cite{QuantumControlTexts}.
At the same time, there are interesting proposals on quantum circuits
whose elementary operations are composed by adiabatic processes%
~\cite{AdiabaticControl}.
The reason why the adiabatic processes are employed is that
the operations governed by the adiabatic processes are 
expected to be stable.
%
On the other hand, the manipulation that involves the phase anholonomy
is expected to be stable under the perturbation, due to its topological 
nature.

Our scheme proposed here also relies on the adiabatic processes and 
employs nonconventional, Cheon's anholonomies in quantum maps.
The aim is to evolve a quantum state (``the initial target'') 
into another state (``the final target'').
What we need to carry it out is twofold:
One is an ``unperturbed'' Hamiltonian $\hat{H}_0$, whose eigenstates
must contain the two target states.
The other is a normalized vector $\ket{v}$, which must have
nonzero overlappings between the two target states.
Under the influence of a periodically pulsed perturbation 
$\hat{V}=\ketbra{v}{v}$ with its period $T$ and strength $\lambda$,
the system is described 
by the kicked Hamiltonian $\hat{H}(t)$~(\ref{eq:kickedHamiltonian}).
The use of the quasienergy anholonomy of 
the corresponding Floquet operator $\hat{U}_{\lambda}$~(\ref{eq:defUlambda})
is straightforward 
if  $\hat{H}_0$ is bounded and contains only discrete eigenenergies
and $T$ is smaller than $2\pi\hbar/W$, where
$W$ is the difference between the maximum and the minimum eigenenergies
of $\hat{H}_0$.
Otherwise, we need to achieve these conditions effectively, 
by adjusting $\ket{v}$.
For example, $\ket{v}$ needs to be prepared to have no overlapping 
with the eigenstates that have higher eigenergies to make
an effective energy cutoff on $\hat{H}_0$.

Once we prepare such $\hat{U}_{\lambda}$, it is straightforward to
realize the manipulation, at least, in theory.
To convert a state vector, which is initially in
an eigenstate of $\hat{U}_0$, to the nearest higher eigenstate 
of $\hat{U}_0$, is achieved by applying the periodically pulsed perturbation 
$\hat{V}=\ketbra{v}{v}$, 
whose strength $\lambda$ is adiabatically increased from 0 to $2\pi$. 
Note that at the final stage of the manipulation, we may switch off
the perturbation suddenly, due to the periodicity of the 
Floquet operator $\hat{U}_{2\pi} = \hat{U}_{0}$.
This closes a ``cycle.''
By repeating the cycle, the final state can be 
any eigenstate of $\hat{U}_0$.
Note that, along the operation, $\ket{v}$ may vary adiabatically. 
In other words, the adiabatically slow fluctuation on $\ket{v}$ 
does no harm.
We remark that an application of the present procedure to anholonomic
adiabatic quantum computation is described in a separate 
publication~\cite{TM061}.

The strongest limitations of the present scheme,
in our opinion, is that the target states for the manipulation
must be eigenstates of $\hat{H}_0$. 
Superpositions of the eigenstates of $\hat{H}_0$ cannot be the targets
due to the presence of dynamical phases that generally diverge in adiabatic
processes.
Note that, however, there is no obstacle to 
handle ``superposed states'' when they are eigenstates of $\hat{H}_0$.
Furthermore, if we could introduce Cheon's anholonomies to
the systems whose quasienergy are degenerate, it may be possible to
carry out a coherent manipulation within a degenerate eigenspace.
This motivates us to seek an extension of the eigenspace anholonomy 
for degenerate eigenspace, 
i.e., Cheon's anholonomies \'a la Wilczek and Zee.

\section{Discussion and summary}
\label{sec:summary}

We have discussed Cheon's anholonomies in a family of quantum 
map~(\ref{eq:expandUlambda}) with a rank-one projection $\hat{V}$.
Although our geometrical argument in Section~\ref{sec:abundance} assures
the abundance of the systems that exhibit the anholonomies,
we still do not have any systematic way to find such systems,
except the quantum map~(\ref{eq:expandUlambda}).
In order to suggest exploring other examples of the anholonomies,
we summarize conditions to find the anholonomies.
Two ingredients in our Floquet operator~(\ref{eq:expandUlambda})
facilitate us to find the anholonomies:
(a) the periodicity of the Floquet operator for the parameter $\lambda$ 
enforces the periodicity of the spectrum;
and (b) the positivity of the perturbation assures the monotonic increment
of each quasienergy for the increment of $\lambda$.
These two facts imply that $E_n(\lambda)$ arrives at a higher excited 
quasienergy $E_{n + \delta n}(\lambda)$ ($\delta n>0$) 
after an increment of $\lambda$ by the period $2\pi$.
To realize the first condition, $\hat{V}$ needs not to be a 
projection operator. For the Floquet operator~(\ref{eq:defUlambda}),
the condition for the periodicity is 
$e^{-i \Lambda\hat{V}/\hbar} = 1$, where $\Lambda$ is the period.
In terms of the eigenvalues $\set{v_n}_n$ of $\hat{V}$, this condition 
is that $\Lambda v_n / (2\pi\hbar)$ is an integer for all $n$.
Although we suppose that the anholonomies may be realized without 
the condition (b), we are not aware of any examples,
except the trivial cases, e.g., $\hat{V}$ is negative definite.
Furthermore, the above two conditions generally do not determine the exact
value of $\delta n$, which is the increment of the quantum number 
after a single cycle,
whereas $\delta n=1$ for a rank-1 projection $\hat{V}$ is
shown in Section~\ref{sec:existance}.
The value of $\delta n$ could be determined by the geometric
argument shown in Section~\ref{sec:abundance}.
However, no systematic algorithm to compute $\delta n$ 
from a given family of Floquet operators is known to us.

\begin{acknowledgments}
  M.M. would like to thank Professor I. Ohba and Professor H. Nakazato
  for useful comments.
  A.T. wishes to thank Professor A.~Shudo, Professor K.~Nemoto, and
  Professor M.~Murao for useful conversations.
\end{acknowledgments}

\appendix

\section{A reduction of Hilbert space for 
  a quantum map under a rank-$1$ perturbation}
\label{sec:normalForm}

We explain a procedure to reduce the Hilbert space for 
the quantum map under a rank-$1$ perturbation~(\ref{eq:expandUlambda}).
Let us start from $\lambda = \lambda_0$.
Assume that $\hat{U}_{\lambda_0}$ has a pure point spectrum
(i.e., the eigenvectors of $\hat{U}_{\lambda}$ 
form a complete orthogonal system)~\cite{KatoPurePoint}.
We exclude the case that $\ket{v}$ is an eigenvector of $\hat{U}_{\lambda_0}$
because this implies that the whole Hilbert space becomes trivial, 
as is explained in Section~\ref{sec:rank1}.
An eigenspace $\Hilbert_z$ of $\hat{U}_{\lambda_0}$, 
where $z$ is the corresponding eigenvalue,
is reduced as follows.
First, we introduce $\Hilbert_z^{\rm n}$, which is 
a subspace of $\Hilbert_z$ and orthogonal to $\ket{v}$:
\begin{equation}
  \Hilbert_z^{\rm n}\equiv {\rm span}
  \set{\ket{\xi}\in\Hilbert_z; \bracket{v}{\xi}=0}.
\end{equation}
We exclude $\Hilbert_z^{\rm n}$, since this is a trivial eigenspace 
of $\hat{U}_{\lambda}$ (see Eq.~(\ref{eq:orthoTriviall})).
If the remainder $\Hilbert_z^{\rm p} \equiv
\Hilbert_z\ominus\Hilbert_z^{\rm n}$ is not $\set{0}$,
$\Hilbert_z^{\rm p}$ is a one-dimensional eigenspace of 
$\hat{U}_{\lambda_0}$. Hence the degeneracy in the eigenvalue $z$ is 
removed.
Then we examine the spectrum of $\hat{U}_{\lambda}$
on the resultant Hilbert space $\Hilbert\equiv\oplus_z \Hilbert_z^{\rm p}$.
On $\Hilbert$, $\hat{U}_{\lambda_0}$ has a pure point and nondegenerate
spectrum. At the same time, all of the eigenvector $\ket{\xi}$ 
of $\hat{U}_{\lambda_0}$ satisfies 
\begin{equation}
  \label{eq:v_phi}
  0 \lvertneqq |\bracket{v}{\xi}| \lvertneqq 1.
\end{equation}
For general $\lambda$, we assume that $\hat{U}_{\lambda}$ also has
a pure point spectrum.
Namely, we exclude the case that $\hat{U}_{\lambda}$ has
a continuous spectrum, which emerges under some combinations 
of $\hat{U}_{\lambda_0}$ and $\ket{v}$ in an infinite dimensional 
$\Hilbert$~\cite{Combescure:JSP-59-679}.
As a result of the reduction of $\Hilbert$, 
any eigenvector $\ket{\xi}$ of $\hat{U}_{\lambda}$
also satisfies the inequality~(\ref{eq:v_phi}).
Hence we proved the inequality~(\ref{eq:v_xi}) in the main text.

\section{Cyclicity}
\label{sec:cyclicity}
When a Hilbert space 
$\overline{{\rm span} \{\hat{U}^m \ket{v}\}_{m=0}^{\infty}}$,
which is induced by a vector $\ket{v}$ and an operator $\hat{U}$,
agrees with the whole Hilbert space,
$\ket{v}$ is called a cyclic vector of $\hat{U}$~\cite{RSICyclicity}.
The notion of the cyclicity is useful to discuss how we choose $\ket{v}$
in the quantum map~(\ref{eq:expandUlambda}) to find Cheon's anholonomies, 
as is explained in Sections~\ref{sec:rank1} to~\ref{sec:examples}.
Hence a review of the cyclicity is shown below, where we assume
that the spectrum of $\hat{U}$ has only discrete and finite components.

A characterization of the cyclic vector $\ket{v}$ for $\hat{U}$
is explained:
Any (normalizable) eigenvector $\ket{\xi}$
of $\hat{U}$ satisfies $\bracket{\xi}{v}\ne0$.
To show this, let $z$ be the eigenvalue corresponding to $\ket{\xi}$.
Due to the cyclicity, $\ket{\xi}$ is a linear
combination of $\{\hat{U}^m \ket{v}\}_{m=0}^{\infty}$,
i.e., $\ket{\xi}
= \sum_{m=0}^{\infty} c_m \hat{U}^m \ket{v}$ 
with appropriate coefficients $c_m$. 
Hence we have
$\bracket{\xi}{\xi} 
= \sum_{m=0}^{\infty} c_m \bra{\xi} \hat{U}^m \ket{v}
= \sum_{m=0}^{\infty} c_m z^m \bracket{\xi}{v}
= \left(\sum_{m=0}^{\infty} c_m z^m\right) \bracket{\xi}{v}$.
Since $\bracket{\xi}{\xi}$ is nonzero and 
$\sum_{m=0}^{\infty} c_m z^m$ is finite, 
we conclude $\bracket{\xi}{v}\ne 0$.
Note that this just proves the fact that the condition (ii') implies
the condition (ii) in Section~\ref{sec:rank1}.

Next, we show that the inverse holds, 
i.e., the condition (ii) implies the condition (ii')
when the spectrum of $\hat{U}$ is nondegenerate.
More precisely, when all eigenvectors $\ket{\xi}$ of $\hat{U}$
satisfy $\bracket{v}{\xi}\ne 0$, 
$\ket{v}$ is a cyclic vector for $\hat{U}$.
To show this, we prove that $\set{\hat{U}^m\ket{v}}_{m=0}^{N-1}$
are linearly independent, where $N$ is the number of the eigenvalues.
Namely, for an $N$-dimensional vector 
$\bvec{c} = (c_0, c_1, \ldots, c_{N-1})$, we show that
\begin{equation}
  \label{eq:vmIndependence}
  \sum_{m=0}^{N-1} c_m\hat{U}^m\ket{v} = 0
\end{equation}
implies $\bvec{c} = 0$.
Let $z_n$ and $\ket{\xi_n}$ denote an eigenvalue of $\hat{U}$
and the corresponding eigenvector, respectively
($n = 0, 1, \ldots, N-1$).
From Eq.~(\ref{eq:vmIndependence}), we have
$
  \sum_{m=0}^{N-1}\braOket{\xi_n}{\hat{U}^m}{v} c_m 
  = \bracket{\xi_n}{v}\sum_{m=0}^{N-1}(z_n)^m c_m = 0.
$
The assumption $\bracket{\xi_n}{v}\ne 0$ implies 
$\sum_{m=0}^{N-1}(z_n)^m c_m =0$ for all $n$.
This is written as $A\bvec{c}=0$, where $A$ is the 
$N$-dimensional square matrix whose $(m, n)$-element is $(z_n)^m$.
Accordingly we encounter a Vandermonde determinant
\begin{align}
  \det A 
  &=
  \begin{vmatrix}
    1& z_0& (z_0)^2& \dots& (z_0)^{N-1}\\
    1& z_1& (z_1)^2& \dots& (z_1)^{N-1}\\
    \hdotsfor{5}\\
    1& z_{N-1}& (z_{N-1})^2& \dots& (z_{N-1})^{N-1}
  \end{vmatrix}
  \nonumber \\
  &=
  \prod_{n''>n'} (z_{n''} - z_{n'})
\end{align}
and we have $\det A \ne 0$ due to the absence of spectrum
degeneracy of $\hat{U}$. Hence we have $\bvec{c} = 0$.

We can now examine the condition when $\hat{U}$ has a cyclic vector
to see the correspondence between the conditions (i) and (i') in
Section~\ref{sec:rank1}.
If the spectrum of $\hat{U}$ is nondegenerate, $\hat{U}$ has 
a cyclic vector, e.g., $\sum_{n=0}^{N-1} \ket{\xi_n}$, 
from the above discussion.
Furthermore, $\hat{U}$ has a cyclic vector only when $\hat{U}$ has
no degenerate eigenvalue. To show the latter,
we examine its contraposition. Hence we assume that 
$\hat{U}$ has a degenerate eigenvalue, i.e., $\det A = 0$.
Accordingly we have a nonzero $\bvec{c}$ 
that satisfies $A \bvec{c} = \bvec{0}$,
i.e., $\sum_{m=0}^{N-1}(z_n)^m c_m =0$ for all $n$.
With such $\bvec{c}$ and arbitrary $d_n$, we have
$0 
= \sum_{n=0}^{N-1}d_n \bra{\xi_n}\sum_{m=0}^{N-1}(z_n)^m c_m 
= \sum_{n=0}^{N-1}d_n \bra{\xi_n}\sum_{m=0}^{N-1}\hat{U}^m c_m,$
i.e., $\sum_{m=0}^{N-1} c_m \hat{U}^m = 0$.
Accordingly, with any vector $\ket{v}$, we have
$\sum_{m=0}^{N-1} c_m \hat{U}^m \ket{v} = 0$, i.e.,
$\set{\hat{U}^m \ket{v}}_{m=0}^{N-1}$ is linearly dependent
and any $\ket{v}$ cannot be a cyclic vector for $\hat{U}$.
Thus the degenerate eigenvalue of $\hat{U}$ leads to an absence of
its cyclic vector.

To summarize this appendix, we explain the conditions
(i') and (ii') in Section~\ref{sec:rank1}. 
If the spectrum of $\hat{U}$ is finite and nondegenerate, 
$\hat{U}$ has a cyclic vector. Furthermore, if $\ket{v}$ satisfies
$\bracket{v}{\xi}\ne0$ for any eigenvector $\ket{\xi}$ of $\hat{U}$,
$\ket{v}$ is a cyclic vector of $\hat{U}$.


\begin{thebibliography}{99}
\bibitem{endnote:leveldynamics}
  More precisely, we need a set of equations 
  for $\bra{\xi_n(\lambda)}$ to obtain a complete set of the equations 
  of motion~\cite{Nakamura:PRA-35-5294}.%
\bibitem{note:infiniteSpan}
  Since we assume that
  the dimensionality of the whole Hilbert space is finite (say, $N$),
  $\{(\hat{U}_0)^m \ket{v}\}_{m=0}^{\infty}$ is linearly dependent.
  We may use an exact truncation 
  $\overline{{\rm span} \{(\hat{U}_0)^m \ket{v}\}_{m=0}^{\infty}}
  ={\rm span} \{(\hat{U}_0)^m \ket{v}\}_{m=0}^{N-1}$
  in the condition of the cyclicity.
\end{thebibliography}



\end{document}